# A NEW CRITERION FOR BAR–FORMING INSTABILITY IN RAPIDLY ROTATING GASEOUS AND STELLAR SYSTEMS. II. NONAXISYMMETRIC FORM

Dimitris M. Christodoulou[1], Isaac Shlosman[2,3], and Joel E. Tohline[4]

## ABSTRACT

We have previously introduced the parameter $\alpha$ as an indicator of stability to $m = 2$ nonaxisymmetric modes in rotating, self–gravitating, axisymmetric, gaseous ($\alpha \lesssim 0.34$) and stellar ($\alpha \lesssim 0.25$) systems. This parameter can be written as $\alpha = (ft/2)^{1/2}$, where $t \equiv T/|W|$, $T$ is the total rotational kinetic energy, $W$ is the total gravitational potential energy, and $f$ is a function characteristic of the topology/connectedness and the geometric shape of a system. In this paper, we extend the stability criterion to nonaxisymmetric equilibrium systems by determining empirically the appropriate form of the function $f$ for ellipsoids and elliptical disks and cylinders.

For oblate–like ellipsoidal systems, we find that

$$f = \frac{2\sqrt{1-\eta^2}}{\eta^2}\Big[1 - \frac{E(\sin^{-1}e, \eta^2/e^2)}{F(\sin^{-1}e, \eta^2/e^2)}\Big],$$

where $e$ is the meridional eccentricity, $\eta$ is the equatorial eccentricity, and $F$ and $E$ are the incomplete elliptic integrals of the first and second kind, respectively, with amplitude $\sin^{-1} e$ and parameter $\eta^2/e^2$. For prolate–like ellipsoidal systems, we find an analogous expression that reduces to $f \equiv 0$ in the limiting case of infinite cylinders.

We test the validity of this extension of the stability indicator $\alpha$ by considering its predictions for previously published, gaseous and stellar, nonaxisymmetric models. The above formulation and critical values account accurately for the stability properties of $m = 2$ modes in gaseous Riemann S–type ellipsoids (including the Jacobi and Dedekind ellipsoids) and elliptical Riemann disks as well as in stellar elliptical Freeman disks and cylinders: all these systems are dynamically stable except the stellar elliptical Freeman disks that exhibit a relatively small region of $m = 2$ dynamical instability. A partial disagreement in the case of stellar Freeman ellipsoids in maximum rotation may be due to that the region of instability has not been previously determined with sufficient accuracy.

[1]Virginia Institute for Theoretical Astronomy, Department of Astronomy, University of Virginia, P.O. Box 3818, Charlottesville, VA 22903

[2]Department of Physics & Astronomy, University of Kentucky, Lexington, KY 40506

[3]Gauss Foundation Fellow

[4]Department of Physics & Astronomy, Louisiana State University, Baton Rouge, LA 70803

*Subject headings:* galaxies: evolution – galaxies: structure – hydrodynamics – instabilities – stars: formation

## 1    Introduction

In a recent paper (Christodoulou, Shlosman, & Tohline 1994; hereafter referred to as Paper I), we proposed a new criterion for stability of rotating, self–gravitating, axisymmetric systems to $m = 2$ perturbations. The criterion was formulated in terms of the angular momentum rather than the energy content of a system and was expressed by the conditions $\alpha \lesssim 0.25$–$0.26$ and $\alpha \lesssim 0.34$–$0.35$ for stability of stellar and gaseous systems, respectively. The stability indicator $\alpha$ was defined for uniformly rotating systems by

$$\alpha \equiv \frac{t}{\chi} \equiv \sqrt{\frac{1}{2}ft}, \qquad (1.1)$$

where $t = T/|W|$ is the ratio of the rotational kinetic energy to the absolute value of the gravitational potential energy, $\chi = \Omega/\Omega_J$ is the ratio of the rotation frequency to the Jeans frequency introduced by self–gravity, and $f$ is a function that depends on both the geometry and the topology of a system.

The first expression of $\alpha$ in equation (1.1) is inconvenient for applications because the ratio $\chi = \Omega/\Omega_J$ cannot be determined easily in all cases of interest. For example, in differentially rotating systems, it is not known how a weighted value of $\Omega$ should be obtained. In centrally condensed and/or nonaxisymmetric systems, it is not known how a weighted value of $\Omega_J$ should be estimated. The second expression of $\alpha$ in equation (1.1) is not plagued by these difficulties but the functional form of the term $f$ must be known for applications to systems with various geometrical/topological structures. A typical case where $f$ is not generally known is that of systems with multiply–connected regions (see e.g. the toroidal models and the Toomre–Zang disks in Paper I).

In Paper I, $f$ was determined for homogeneous, uniformly rotating, oblate spheroidal systems as

$$f = \frac{A_1(e)}{\sqrt{1-e^2}} \frac{e}{\sin^{-1} e}, \qquad (1.2)$$

where $e$ is the meridional eccentricity and

$$A_1(e) = \frac{\sqrt{1-e^2}}{e^3} \sin^{-1} e - \frac{1-e^2}{e^2}. \qquad (1.3)$$

In particular, $f = 1$ for disks and $f = 2/3$ for spheres. It was also demonstrated in Paper I that the second expression of $\alpha$ in equation (1.1) with $f$ given by equation (1.2) provides an accurate stability indicator for various oblate spheroidal and disk–like models with nonuniform density and differential rotation. Furthermore, the parameter $\alpha$ was found to be more sensitive than the ratio $T/|W|$ that appears in the stability criterion proposed for stellar systems ($T/|W| \lesssim 0.14$) by Ostriker & Peebles (1973; hereafter referred to as OP) and in the analogous criterion for gaseous systems ($T/|W| \lesssim 0.27$).

The success of $\alpha$ as an indicator of stability to $m = 2$ modes in various axisymmetric systems prompted us to attempt its generalization in the case of simply–connected, nonaxisymmetric systems. In §2, we describe an empirical extension of equation (1.2) to oblate–like ellipsoids. With this new definition of $f$, $\alpha$ is still given by the second expression in equation (1.1) and its critical values for marginal stability remain unchanged. In the first four subsections of §3, we discuss the predictions of the parameter $\alpha$ for the stability of several oblate–like gaseous and stellar models (ellipsoids and elliptical disks) that have been previously constructed and studied by various researchers. In §3.5, we discuss prolate–like geometries and the predictions of $\alpha$ for infinite elliptical cylinders. In §4, we summarize our conclusions.

## 2  The parameters $\alpha$ and $f$ for Oblate–Like Ellipsoids

Motivated by the results described in Paper I, we adopt the expression

$$\alpha \equiv \sqrt{\frac{1}{2}ft}, \tag{2.1}$$

as a general definition of the parameter $\alpha$ and we obtain a general expression for the function $f$ valid for oblate–like ellipsoids. This method avoids the use of the equations $\alpha = t/\chi$ and $\chi = \Omega/\Omega_J$ (see §1 above). This is convenient since it is not clear what form should be adopted for the Jeans frequency $\Omega_J$ in nonaxisymmetric systems.

We notice that the terms $A_1(e)/(1-e^2)^{1/2}$ and $e/\sin^{-1} e$ in equation (1.2) have been introduced by the gravitational potential $\Phi$ and the gravitational potential energy $W$, respectively (Paper I). For oblate–like ellipsoids with three unequal axes $a > b > c$ rotating about the short axis $c$, $A_1$ is a function of both the meridional eccentricity $e = (1 - c^2/a^2)^{1/2}$ and the equatorial eccentricity $\eta = (1 - b^2/a^2)^{1/2}$ (Chandrasekhar 1969; hereafter referred to as EFE), i.e.

$$A_1(e,\eta) = \frac{2\sqrt{1-e^2}\sqrt{1-\eta^2}}{e\eta^2}\Big[F(\sin^{-1} e, \eta^2/e^2) - E(\sin^{-1} e, \eta^2/e^2)\Big], \tag{2.2}$$

where $F$ and $E$ are the incomplete elliptic integrals of the first and second kind, respectively, and

$$F(\sin^{-1} e, \eta^2/e^2) \equiv \int_0^{\sin^{-1} e} \frac{d\phi}{\sqrt{1 - (\eta^2/e^2)\sin^2 \phi}}, \tag{2.3}$$

$$E(\sin^{-1} e, \eta^2/e^2) \equiv \int_0^{\sin^{-1} e} \sqrt{1 - (\eta^2/e^2)\sin^2 \phi}\, d\phi. \tag{2.4}$$

In definitions (2.3) and (2.4), the upper limit of the integrals $\sin^{-1} e$ is called the amplitude while the term $\eta^2/e^2$ is called the parameter (Abramowitz & Stegun 1972).

In addition, $W \propto F(\sin^{-1} e, \eta^2/e^2)/e$ in oblate–like ellipsoids (e.g. Freeman 1966b) and $W \propto F(\pi/2, \eta^2)$ in elliptical disks (e.g. Freeman 1966c; Weinberg & Tremaine 1983). Combining the above results, we assume that equation (1.2) generalizes as

$$f = \frac{A_1(e,\eta)}{\sqrt{1-e^2}} \frac{e}{F(\sin^{-1} e, \eta^2/e^2)}, \qquad (2.5)$$

where $A_1(e,\eta)$ is given by equation (2.2). For applications, we adopt the expression

$$f = \frac{2\sqrt{1-\eta^2}}{\eta^2}\Big[1 - \frac{E(\sin^{-1} e, \eta^2/e^2)}{F(\sin^{-1} e, \eta^2/e^2)}\Big], \qquad (2.6)$$

which is derived by combining equations (2.2) and (2.5). For a given model of known $t = T/|W|$, we determine the stability parameter $\alpha$ from equations (2.1) and (2.6) and, as in Paper I, we adopt $\alpha \lesssim 0.25$–$0.26$ and $\alpha \lesssim 0.34$–$0.35$ to indicate stability to $m = 2$ modes in stellar and gaseous systems, respectively.

FIGURE 1. The function $f$ is plotted versus equatorial eccentricity $\eta$ for elliptical disks ($e = 1$) and for Jacobi ellipsoids (EFE).

As we shall see below, equation (2.6) resolves a well–known discrepancy in the case of "needles" where $e, \eta \to 1$. Such objects are stable to $m = 2$ perturbations (e.g. EFE; Tremaine 1976; Weinberg 1983). In the limit $e, \eta \to 1$, $t \to 1/2$ and thus any criterion based on $t$ alone fails to predict stability. On the other hand, equation (2.6) shows that $f \to 0$ in this limit and, thus, equation (2.1) predicts stability to $m = 2$ modes. This behavior is seen in Figure 1 where the function $f(\eta)$ is plotted for elliptical disks ($e = 1$) and for Jacobi ellipsoids in which $e$ and $\eta$ are uniquely related to each other (e.g. EFE; Christodoulou, Kazanas, Shlosman, & Tohline 1994, hereafter referred to as CKST).

For axisymmetric systems with $\eta = 0$, the above equations reduce to the expressions given in Paper I. For nonaxisymmetric disks, $e = 1$ and the elliptic integrals above become complete with amplitude $\pi/2$ and parameter $\eta^2$ (Abramowitz & Stegun 1972; Weinberg & Tremaine 1983). Finally, the corresponding expression for the "weighted" Jeans frequency $\Omega_J$ can be determined from the second equality in equation (1.1) and equation (2.6). However, such an expression is of limited use since $\Omega_J$ turns out to be a complicated function of $\eta$ mainly because of the complexity in the functional form of the kinetic energy $T$ in stellar systems.

Analogous equations valid for prolate–like ellipsoids are discussed in §3.5 below in conjunction with infinite elliptical cylinders.

FIGURE 2. *Contours of the ratio $t = T/|W|$ for gaseous Riemann S–type ellipsoids are plotted in the $(x, b/a)$ plane. All these objects are stable to second–harmonic perurbations (EFE). In the needle limit $b/a \to 0$, $t \to 1/2$ and is thus insensitive to the dynamics of stable strongly nonaxisymmetric Riemann ellipsoids.*

### 3 Previously Studied Nonaxisymmetric Systems

#### 3.1 Gaseous Riemann S–type Ellipsoids

These uniformly rotating incompressible models are studied in detail in EFE. A brief description has recently been given also by CKST. The ellipsoidal figures are oblate–like in shape with axes $a > b > c$ and the rotation takes place about the $c$ axis with frequency $\Omega$.

FIGURE 3. Contours of the parameter $\alpha$ for gaseous Riemann S–type ellipsoids are plotted in the $(x, b/a)$ plane. All these objects are stable to second–harmonic perurbations (EFE). This result is confirmed by the parameter $\alpha$ since $\alpha \lesssim 0.34$ everywhere. In the needle limit $b/a \to 0$, $\alpha \to 0$ showing its sensitivity to the dynamics of stable strongly nonaxisymmetric Riemann ellipsoids.

All gaseous Riemann S–type ellipsoids, including the Jacobi and Dedekind ellipsoids, are stable to second–harmonic perturbations. This behavior is not captured by the stability criterion $t = T/|W| \lesssim 0.27$ that appears to predict instability in the needle limit $e, \eta \to 1$ where $t \to 1/2$. The criterion based on the parameter $\alpha$ does not suffer from similar difficulties. The results are illustrated in Figures 2 and 3 where contour plots of the parameters $t$ and $\alpha$ are shown in the $(x, b/a)$ plane. The axis ratio $b/a = (1 - \eta^2)^{1/2}$ while the parameter $x = [ab/(a^2 + b^2)](\zeta/\Omega)$, where $\zeta$ is the vorticity in a frame rotating with frequency $\Omega$ in which the ellipsoidal figures appear to be stationary.

There exists only one point of marginal stability in Figures 2 and 3 at $b/a = 1$, $x = 1$, where $t = 0.2738$ and $\alpha = 0.3410$ (see also Paper I and CKST). This point denotes the onset of second–harmonic dynamical instability in Maclaurin spheroids. All the remaining points represent stable Riemann S–type ellipsoids including the stable Maclaurin spheroids ($b/a = 1$). (In fact, all sequences of Riemann S–type ellipsoids characterized by different values of $x$ or $\zeta/\Omega$ bifurcate from the stable part of the Maclaurin sequence.) The parameter $\alpha$, in the form adopted in §2 above, confirms the stability of all these objects to second–harmonic perturbations since $\alpha \lesssim 0.34$ everywhere — even in the region $|x| > 5$ not covered by Figure 3.

### 3.2  Gaseous Riemann Disks

The structure and secular evolution of these uniformly rotating compressible models with equatorial axes $a \geq b$ have been studied by Weinberg & Tremaine (1983). The compressible Riemann disks and the incompressible ellipsoids of §3.1 exhibit very similar dynamical properties. In particular, all equilibrium sequences bifurcate from the stable part of the corresponding Maclaurin sequence of circular disks and all objects are stable to $m = 2$ perturbations (Weinberg 1983).

FIGURE 4. *Contours of the ratio $t = T/|W|$ for gaseous Riemann disks are plotted in the $(\zeta/\Omega, b/a)$ plane. All these objects are stable to $m = 2$ perurbations (Weinberg 1983). In the needle limit $b/a \to 0$, $t \to 1/2$ and is thus insensitive to the dynamics of stable strongly nonaxisymmetric Riemann disks.*

Despite a coordinate change from $x$ in ellipsoids to $\zeta/\Omega$ in disks (that is done to avoid presenting nearly identical diagrams to Figures 2 and 3 and to depict the region near $x = \zeta/\Omega = 0$ in more detail), the dynamical similarities between Riemann disks and ellipsoids are apparent in a comparison of the behavior of $t = T/|W|$ between Figure 2 above and Figure 4 that shows a contour plot of $t$ for disks in the $(\zeta/\Omega, b/a)$ plane. An analogous contour plot of $t$ in the $(b/a, \Omega^2)$ plane, separated into two figures, can also be found in Weinberg & Tremaine (1983).

FIGURE 5. *Contours of the parameter $\alpha$ for gaseous Riemann disks are plotted in the $(\zeta/\Omega, b/a)$ plane. All these objects are stable to $m = 2$ perurbations (Weinberg 1983). This result is confirmed by the parameter $\alpha$ since $\alpha \lesssim 0.35$ everywhere. In the needle limit $b/a \to 0$, $\alpha \to 0$ showing its sensitivity to the dynamics of stable strongly nonaxisymmetric Riemann disks.*

In the needle limit $\eta \to 1$ ($b/a \to 0$), the ratio $t = T/|W| \to 1/2$ and the stability criterion $t \lesssim 0.27$ would predict again instability. In contrast, the parameter $\alpha$, shown in Figure 5 in the $(\zeta/\Omega, b/a)$ plane, confirms the stability of all Riemann disks to $m = 2$ modes. There exists only one point of marginal stability in Figures 4 and 5 at $b/a = 1$, $\zeta/\Omega = 2$, where $t = 1/4$ and $\alpha = 0.3536$. In exact analogy to the ellipsoids of §3.1, this point denotes the onset of $m = 2$ dynamical instability in circular Maclaurin disks (see Paper I for more details). All the other points in Figure 5 represent dynamically stable disks since $\alpha \lesssim 0.35$. This is true also for the disks lying in the region $|\zeta/\Omega| > 5$ not covered by Figure 5.

*3.3   Stellar Elliptical Freeman Disks*

These models were constructed by Freeman (1966c) and their stability to $m = 2$ modes was investigated by Tremaine (1976). Brief discussions of the stability properties of Freeman disks can also be found in Hunter (1974) and Fridman & Polyachenko (1984). Following Hunter (1974), we plot in Figure 6 contours of the ratio $t = T/|W|$ in the $(\chi^2, b/a)$ plane. Here, $a \geq b$, $\chi = \Omega/\Omega_J$, $\Omega$ is the (mean) rotation frequency of the coordinate frame, and $\Omega_J$ is the equatorial Jeans frequency that depends on the function $A_1(e = 1, \eta)$.

This definition of $\Omega$ in stellar systems with circulation superimposed to the mean rotation (Freeman 1966b,c; Hunter 1974) is not equivalent to the definition adopted in §3.1 for gaseous systems (see Paper I). The difference is also reflected in the behavior of $t$ in the two types of systems (CKST). For example, it explains why the Dedekind sequence bifurcates from the Maclaurin sequence at $t = 0$ in stellar systems but at $t=0.125$ and $t=0.1375$ in gaseous disks and spheroids, respectively.

FIGURE 6. *Contours of the ratio $t = T/|W|$ for stellar Freeman disks are plotted in the $(\chi^2, b/a)$ plane. A roughly triangular region of $m = 2$ dynamical instability at the upper corner of the diagram is marked approximately by the $t = 0.14$ contour (Tremaine 1976). As in the gaseous nonrotating Maclaurin disk ($\zeta/\Omega = -2$, $b/a = 1$ in Figure 4), $t = 0$ for the $\Omega$–model with $b/a = 1$ and $\chi^2 = 5/8$ (cf. Kalnajs 1972). Unlike in gaseous disks, $t \to 0$ also in the Dedekind limit $\chi \to 0$ implying that different definitions of $t$ and $\Omega$ are used in the two cases (CKST; Paper I).*

In Figure 6, the segment $b/a = 1$, $5/8 \leq \chi^2 \leq 1$ ($0 \leq T/|W| \leq 1/2$) describes the axisymmetric $\Omega$–models studied by Kalnajs (1972) and Kalnajs & Athanassoula (1974). Note that the dimensionless rotation frequency $\chi_K$ that appears in these works (and in §2.3 of Paper I) is related to the $\chi$ used here by $\chi_K = \chi(8\chi^2-5)/3$; for details see Tremaine (1976). The segment $b/a = 1$, $0 \leq \chi^2 \leq 5/24$ ($0 \leq T/|W| \leq 0.1286$) describes slowly rotating circular stellar disks that are secularly unstable because of slow mass loss (hence also slow angular momentum loss). These circular disks evolve toward the "stellar–disk Dedekind sequence" $\chi = 0$ where they finally relax on the segment $0.4 \lesssim b/a \leq 1$ becoming nonaxisymmetric and nonrotating (see Hunter 1974 and CKST). In the final state, the persisting azimuthal flow is interpreted as circulation (Freeman 1966c), hence it does not contribute to the mean rotational kinetic energy resulting in $T = 0$ and $t = 0$.

FIGURE 7. *Contours of the parameter $\alpha$ for stellar Freeman disks are plotted in the $(\chi^2, b/a)$ plane. The roughly triangular region of $m = 2$ dynamical instability at the upper corner of the diagram (Tremaine 1976) is now marked approximately by the $\alpha = 0.25$ contour. Notice that $\alpha \to 0$ in the needle limit $b/a \to 0$, in the Dedekind limit $\chi \to 0$, and for the $\Omega$–model with $b/a = 1$ and $\chi^2 = 5/8$ (cf. Kalnajs 1972).*

Tremaine (1976) has discovered that $m = 2$ dynamical instability appears only in a roughly triangular region at the upper right corner in Figure 6. The $\Omega$–models of Kalnajs (1972) with $b/a = 1$, $5/6 \leq \chi^2 \leq 1$ ($0.1286 \leq T/|W| \leq 1/2$) mark the upper boundary of the unstable region. This region is also bounded approximately by the $T/|W| = 0.14$ contour and

by the segment $0.7296 \leq b/a \leq 1$, $\chi^2 = 1$. The point of marginal stability at $b/a = 0.7296$, $\chi^2 = 1$ is in very good agreement with the point $a/b = 1.3707$ of marginal stability in gaseous "disk–like Riemann ellipsoids of type I" (EFE). This coincidence is discussed in detail by Tremaine (1976). It occurs because in the maximum–rotation limit $\chi = 1$ the properties of the two types of systems and the corresponding definitions of $\Omega$ become identical.

A contour plot of the parameter $\alpha$ in the $(\chi^2, b/a)$ plane is shown in Figure 7. The triangular region of $m = 2$ dynamical instability at the upper right corner of this plot is now bounded approximately by the $\alpha = 0.25$ contour. The agreement between this value and the contour value $t = 0.14$ in Figure 6 is not just an expected coincidence due to the use of equation (2.1) with $e=1$ and $f(\eta) \approx 1$ (see Tables 1 and 2 in Paper I). We have compared the variations of the two parameters along the marginal stability line given by Tremaine (1976). The variation of $\alpha$ along this line is significantly smaller than that of $t$. Specifically, Tremaine (1976) finds that $t$ varies between 0.1286 (at $b/a = 1$, $\chi^2 = 5/6$) and 0.1446 (at $b/a = 0.7296$, $\chi^2 = 1$). These values correspond to a variation in $t$ of about 11%–12%. We find that $\alpha$ varies between 0.2536 and 0.2385 at the corresponding end–points. These values produce a variation in $\alpha$ of about 6%. The comparison indicates that, as in the axisymmetric systems of Paper I, the parameter $\alpha$ is more sensitive than $t$ as a stability indicator of $m = 2$ modes in elliptical disks.

Furthermore, the contour plot of $\alpha$ (Figure 7) confirms the stability of needles unlike the ratio $T/|W|$ of the OP criterion (Figure 6). Notice, in particular, that $\alpha \to 0$ smoothly in the needle limit $b/a \to 0$ as well as in the Dedekind limit $\chi \to 0$. Finally, there exist two points of marginal stability with $\alpha = 0.2536$ along the line $b/a = 1$ in Figure 7. The corresponding critical values are $\chi^2 = 5/24$ and $\chi^2 = 5/6$. As was described above and in CKST, the segment $\chi^2 \leq 5/24$ with $t \leq 0.1286$ denotes the appearance of secular instability in circular disks unlike the segment $\chi^2 \geq 5/6$ with $t \geq 0.1286$ which denotes the appearance of dynamical instability (see also Hunter 1974).

### 3.4  Stellar Freeman Ellipsoids

These models of stellar homogeneous ellipsoids were constructed by Freeman (1966b). The Freeman ellipsoids are "balanced" in the sense that the gravitational and centrifugal forces are by assumption exactly equal to each other along the major axis. This ensures that $\chi = 1$, where $\chi$ is defined as in §3.3 for stellar disks. The stability of these models to $m = 2$ modes is not known with certainty and should be further investigated. Fridman & Polyachenko (1984; hereafter referred to as FP) describe an attempt to locate regions of dynamical instability in the $(b/a, c/a)$ plane, where $a > b, c$ are the principal axes and $c$ is the rotation axis of the ellipsoids. Some of their results are suspect because the marginal stability curve is jagged.

To the extent that balanced Freeman ellipsoids share some common properties with $\chi = 1$ Freeman disks in the regime $c/a << 1$, we expect that flattened ellipsoids with $b/a \approx 1$ should be dynamically unstable to $m = 2$ modes (cf. the unstable disks with $0.7296 \lesssim b/a \leq 1$ in §3.3). This expectation is generally consistent with the results shown in Figure 48 of FP but provides no justification for an additional extension of the unstable region found at intermediate values of $b/a$ and $c/a$. It is this extension that appears to be suspect.

FIGURE 8. *Contours of the ratio $t = T/|W|$ for balanced ($\chi = 1$) Freeman ellipsoids are plotted in the $(b/a, c/a)$ plane. The boundary of a vertical band of $m = 2$ dynamical instability with $b/a \gtrsim 0.73 - 0.90$ (FP) lies between the $t = 0.14$ and $t = 0.20$ contours on the right–hand side of the diagram. (Thus, the OP criterion is not very successful in tracking this boundary.) An extension of the unstable region to intermediate $b/a$, $c/a$ values shown in FP corresponds to low t–values. If real, this extension cannot be understood in terms of the OP stability criterion. For disks ($c = 0$), the exact marginal stability point is located at $b/a = 0.7296$ where $t = 0.1446$ (§3.3). Virtually all models with $b/a < 0.4$ have $t > 0.14$ although many of them are stable according to FP and all of them are stable according to parameter $\alpha$ (see Figure 9).*

Contour plots of $t = T/|W|$ and $\alpha$ for Freeman ellipsoids are shown in Figures 8 and 9, respectively. In the limiting case $c = 0$, the results reduce to those discussed in §3.3 for Freeman disks in maximum rotation ($\chi = 1$). The critical values of both parameters ($t \approx 0.14$ and $\alpha \approx 0.25$) track the boundary of a vertical band of $m = 2$ instability shown in FP at $b/a \gtrsim 0.73 - 0.90$ (for $c/a = 0 - 1$, respectively) although $\alpha$ is obviously more accurate. None of the parameters predicts an extension of the unstable region to intermediate values of $b/a$ and $c/a$ seen in Figure 48 of FP. Finally, $t$ is unable to confirm the stability of models with $b/a < 0.4$ since $t > 0.14$ over most of this region.

FIGURE 9. *Contours of the parameter $\alpha$ for balanced ($\chi = 1$) Freeman ellipsoids are plotted in the $(b/a, c/a)$ plane. The boundary of a vertical band of $m = 2$ dynamical instability with $b/a \gtrsim 0.73 - 0.90$ (FP) is marked approximately by the $\alpha = 0.25$ contour. (The variation of $\alpha$ along the boundary is about 5% as opposed to 28%–38% for $t$ in Figure 8.) All models on the left of the $\alpha = 0.25$ contour are predicted to be stable to $m = 2$ modes by the parameter $\alpha$ but not by the results in FP that show an extension of the unstable region roughly between the $\alpha = 0.15$ and $\alpha = 0.18$ contours. If real, this extension cannot be understood in terms of the $\alpha$ stability criterion. For disks ($c = 0$), the exact marginal stability point is located at $b/a = 0.7296$ where $\alpha = 0.2385$ (§3.3).*

### 3.5  Stellar Elliptical Freeman Cylinders

These infinite cylindrical models were constructed by Freeman (1966a) and their stability properties were investigated by Nishida & Ishizawa (1977) who found them stable to $m = 2$ modes. Hunter (1974) presented a contour plot of a parameter $t$ equivalent to the ratio $T/|W|$. We have repeated the calculation and have reproduced Hunter's result. The contour plot of $t$ is shown in Figure 10 in the $(\chi^2, c/b)$ plane, where we now assume that the axes $a > b \geq c$, $a \to \infty$, and $a$ is also the rotation axis. We see that the OP criterion, based on Hunter's parameter $t$, suffers from the usual problem of predicting an $m = 2$ instability for a large subset of models with $t > 0.14$.

FIGURE 10. *Contours of Hunter's (1974) parameter t (that is equivalent to $T/|W|$) for stellar Freeman cylinders are plotted in the $(\chi^2, c/b)$ plane. For these infinite cylinders we assume that the rotation is about the a axis, $a \to \infty$, and the equatorial axes $b \geq c$. Although $t > 0.14$ in a substantial part of the diagram, all Freeman cylinders are stable to $m = 2$ modes (Nishida & Ishizawa 1977).*

The formulation of §2 is not valid for prolate–like ellipsoidal and cylindrical models that rotate about the longest axis. For prolate–like ellipsoids with axes $a > b > c$ and rotating about the $a$ axis (EFE), the relevant function is $A_2(e, \eta)$ instead of $A_1(e, \eta)$. This function is associated with the longest equatorial axis $b$ and can be written as

$$A_2(e, \eta) = \frac{2e(1-\eta^2)^{3/2}}{\eta^2(e^2-\eta^2)}\Big[E(\sin^{-1}e, k^2) - \frac{\eta^2}{e^2}\frac{1-e^2}{1-\eta^2}F(\sin^{-1}e, k^2) - \frac{e^2-\eta^2}{e\sqrt{1-\eta^2}}\Big], \quad (3.1)$$

where now $e \equiv (1 - c^2/a^2)^{1/2}$ and $\eta \equiv (1 - c^2/b^2)^{1/2}$ since $a$ is the longest axis and $c$ is the shortest axis. The parameter $k^2$ in the elliptic integrals of equation (3.1) is defined by

$$k^2 \equiv \frac{e^2 - \eta^2}{e^2(1-\eta^2)}. \quad (3.2)$$

By analogy to the assumptions made in §2, we are led to adopt the expression

$$f = \frac{eA_2(e, \eta)}{F(\sin^{-1}e, k^2)}, \quad (3.3)$$

for the function $f(e, \eta)$ in prolate–like ellipsoids. Equation (3.3) is analogous to equation (2.5) above. The absence of a term similar to $(1-e^2)^{1/2}$ of equation (2.5) is due to the change in roles of the axes and is justified by a direct calculation in the case of prolate spheroids (see below).

The limiting case of an infinite elliptical cylinder with $a \to \infty$ is of interest here in relation to the models shown in Figure 10. In this case, $e \to 1$ implying that the amplitude $\sin^{-1} e \to \pi/2$ and the parameter $k^2 \to 1$. Hence, the elliptic integrals in equation (3.1) take the asymptotic values $E \to 1$ and $F \to \infty$ but the function $A_2(e=1, \eta)$ remains finite. By equation (3.3), this asymptotic behavior leads to $f = 0$ for all infinite cylinders. Hence, $\alpha = 0$ by equation (2.1), indicating stability to $m = 2$ modes for all Freeman cylinders as well as for all gaseous infinite cylinders. This last conclusion is in agreement with the result of Chandrasekhar (1981) that circular incompressible infinite cylinders are stable to all nonaxisymmetric perturbations including the $m = 2$ modes.

The special case of prolate spheroidal models may be of some interest despite the fact that a particular class of objects, the homogeneous incompressible prolate Maclaurin spheroids, are not equilibrium figures (Florides & Spyrou 1993). For this type of geometry and for uniform rotation, we assume that the axes $a > b = c$ and that $a$ is the rotation axis and we follow the procedure outlined in Paper I. We find again that $t = \frac{1}{2} f (\Omega/\Omega_J)^2$ and $\alpha = (ft/2)^{1/2}$ but now

$$f = \frac{e A_2(e)}{\ln \sqrt{\frac{1+e}{1-e}}}, \qquad (3.4)$$

where, as in EFE, $e = (1 - c^2/a^2)^{1/2}$ and

$$A_2(e) = \frac{1}{e^2} - \frac{1-e^2}{e^3} \ln \sqrt{\frac{1+e}{1-e}}. \qquad (3.5)$$

Equation (3.4) is analogous to equation (1.2) above. It has no meaning for homogeneous incompressible prolate spheroids but, like equation (1.2), it may prove useful when prolate spheroids — stellar or with nonuniform density and/or differential rotation — are examined for stability to $m = 2$ modes. In the limiting cases of $e = 0$ (sphere) and $e = 1$ (infinite circular cylinder), we recover the expected asymptotic values $f = 2/3$ and $f = 0$, respectively.

## 4  Summary

In §2, we have generalized empirically the criterion derived in Paper I for stability of rotating, self–gravitating, axisymmetric, stellar ($\alpha \lesssim 0.25$–$0.26$) and gaseous ($\alpha \lesssim 0.34$–$0.35$) systems to $m = 2$ modes. The generalized criterion has the same critical values for nonaxisymmetric systems as well. As in the axisymmetric case, the fundamental parameter $\alpha$ is determined from the equation

$$\alpha = \sqrt{\frac{1}{2} ft}, \qquad (4.1)$$

where $t = T/|W|$ is the ratio of the rotational kinetic energy to the absolute value of the gravitational potential energy and $f$ is a function that describes the topological and geometrical structure of a system. For the oblate–like ellipsoids studied in this paper, $f$ is given by the generalized expression

$$f = \frac{2\sqrt{1-\eta^2}}{\eta^2}\Big[1 - \frac{E(\sin^{-1} e, \eta^2/e^2)}{F(\sin^{-1} e, \eta^2/e^2)}\Big], \tag{4.2}$$

where $e$ and $\eta$ are the meridional and equatorial eccentricities and $F$ and $E$ are the incomplete elliptic integrals of the first and second kind with amplitude $\sin^{-1} e$ and parameter $\eta^2/e^2$ (Abramowitz & Stegun 1972).

In §3, we have compared the predictions of the above formulation with the known stability properties of nonaxisymmetric, gaseous and stellar models that have been studied previously by various researchers. The parameter $\alpha$ predicts accurately the stability of ellipsoids and elliptical disks and cylinders. Specifically, gaseous Riemann S–type ellipsoids, including the well–known Jacobi and Dedekind ellipsoids, and gaseous Riemann disks are found to be stable to $m = 2$ modes (in agreement with the results in EFE and in Weinberg 1983). Stellar elliptical disks (Freeman 1966c; Hunter 1974) exhibit only a relatively small region of dynamical instability which lies at the corner of the circular limit $\eta = 0$ and the maximum–rotation limit (in agreement with the results of Tremaine 1976). Finally, stellar elliptical cylinders (Freeman 1966a; Hunter 1974) are found to be stable to $m = 2$ modes (in agreement with the results of Nishida & Ishizawa 1977).

This last prediction of the parameter $\alpha$ extends also to gaseous elliptical cylinders (cf. Chandrasekhar 1981 where circular incompressible infinite cylinders are found to be stable to nonaxisymmetric modes). The reason is that the prediction derives from the condition $f = 0$ (hence $\alpha = 0$) that is valid for all types of infinite cylinders. This result was obtained in the infinite–cylinder limit from the relevant expressions for the function $f$ in prolate spheroids and prolate–like ellipsoids. The formulation has been presented in §3.5.

Little is known about the stability of "balanced" (i.e. maximally rotating) Freeman (1966b) stellar ellipsoids with axes $a > b, c$ (see FP). The stability of these models to $m = 2$ perturbations should be investigated further. A band of instability extending to all values of the axes ratio $c/a$ is shown in Figure 48 of FP in the region $b/a \gtrsim 0.73 - 0.90$. This band is generally consistent with the results of Tremaine (1976) and with the prediction of the parameter $\alpha$ in the disk limit $c \to 0$ (both indicating an $m = 2$ dynamical instability in the region $b/a \gtrsim 0.73$ at maximum rotation). The instability band is predicted accurately for all values of $c/a$ by the parameter $\alpha$ (e.g., at $b/a \gtrsim 0.90$ for $c/a = 1$) but not sufficiently well by the OP stability criterion. Also consistent with the disappearance of the instability in infinite cylinders, the band width shrinks toward zero as $c/a$ increases. However, an additional extension of the unstable region to intermediate values of the axes ratios $b/a$ and $c/a$ shown in FP is not understood in the present context (see Figure 9 above) and needs to be examined in future work.

Acknowledgments

We are grateful to J. Ostriker for helpful discussions and suggestions. We thank C. Hunter, A. Toomre, and P. Vandervoort for useful correspondence. IS is grateful to Gauss Foundation for support and to K. Fricke, Director of Universitäts-Sternwarte Göttingen, for hospitality during a stay in which much of this work has been accomplished. This work was supported in part by NASA grants NAGW–1510, NAGW–2447, NAGW–2376, and NAGW–3839, by NSF grant AST–9008166, and by grants from the San Diego Supercomputer Center, the National Center for Supercomputing Applications, and the University of Kentucky Center for Computational Studies.